\documentclass[pra,twocolumn,showpacs,floatfix]{revtex4}
\usepackage[dvips]{graphicx}%
\usepackage{bm,color}
\usepackage{amsmath,amssymb}
\newcommand{\braket}[2]{\langle #1 | #2 \rangle}
\newcommand{\ketbra}[2]{\ket{#1}\bra{#2}} 
\newcommand{\ket}[1]{\left |  #1 \right \rangle}
\newcommand{\bra}[1]{ \left \langle #1  \right |}
\newcommand{\ave}[1]{ \langle #1  \rangle}
\def \tr{{\textrm {Tr}}}

\begin{document}
\title{Amplification uncertainty relation for probabilistic amplifiers}

\author{Ryo Namiki}
\affiliation{Institute for Quantum Computing and Department of Physics and Astronomy,
University of Waterloo, Waterloo, Ontario, N2L 3G1, Canada}
%

\date{\today}
\begin{abstract} 
Traditionally, quantum amplification limit  refers to the property  of inevitable  noise addition on canonical variables when the field amplitude of an unknown state is linearly transformed through a quantum channel.  Recent theoretical studies  have determined amplification limits for cases of probabilistic quantum channels or general quantum operations by  specifying a set of input states or a  state ensemble. 
 However, it remains open how much excess  noise on canonical variables is unavoidable and whether  there exists a fundamental trade-off relation between the canonical pair in  a  general amplification process. 
 In this paper  
 we present an   uncertainty-product form of amplification limits for general quantum operations by assuming  an input ensemble of Gaussian distributed coherent states.  It can be derived as a straightforward consequence of canonical uncertainty relations   and retrieves basic properties of the traditional amplification limit. In addition, our amplification limit turns out to give a physical limitation on probabilistic reduction of an Einstein-Podolsky-Rosen uncertainty. 
 In this regard,  we find a condition that probabilistic amplifiers can be regarded as  local filtering operations to distill entanglement. This condition establishes a clear benchmark to verify an advantage of non-Gaussian operations beyond Gaussian operations with a feasible input set of coherent states and standard homodyne measurements. 
 
\end{abstract}

\pacs{03.67.Hk, 03.65.Ta, 42.50.Ex, 42.50.Xa} 
\maketitle


%

\section{Introduction}

It is fundamental to ask how an amplification of canonical variables modifies the phase-space distribution of amplified states under the physical constraint due to canonical uncertainty relations. The standard theory to address this question is the \textit{so-called}  amplification uncertainty principle   \cite{amp}. 
 It describes the property of inevitable  noise addition on canonical variables when the field amplitude of an unknown state is linearly transformed through a quantum channel. This traditional form of quantum amplification limits is directly derived from the property of canonical variables, and  gives an important  insight on a wide class of  experiments in quantum optics, quantum information science \cite{{YY86},Cerf00,Cerf00R,{Lind00},Andersen05,rmp-clone,Josse06,Koike06,Sab07,Pooser09,Weed12}, and condensed matter physics \cite{Cle10}.   
Unfortunately,  the linearity of amplification maps assumed in this theory is hardly satisfied in the experiments \cite{ComLR}, although this assumption corresponds to a  covariance property that works as an essential theoretical tool to analyze  a general property of  amplification and related cloning maps \cite{Gut06,{Cer05}}.  
 It is more realistic to consider the performance of  amplifiers in a limited input space. In fact,  one can find a practical limitation  by focusing on a set of input states or an ensemble of input states 
\cite{Namiki11R,Chir13,Pande13}.

There has been a growing interest in implementing  probabilistic amplifiers in order to overcome the standard limitation of the traditional amplification limit  \cite{{Ralph09},Ferreyrol2010,Usuga2010,Xiang2010a,Zavatta2011,Chrzanowski2014}.  In these approaches, one can obtain essentially noiselessly amplified coherent states with a certain probability by conditionally choosing the output of the process.  Recent theoretical studies  have determined amplification limits for such cases of probabilistic quantum channels or general quantum operations \cite{Chir13,Pande13}. 
Certainly, these results can reach beyond the coverage of the traditional theory.  However, it seems difficult to find a precise interrelation between these theories. For example, it is not clear whether  the  traditional form can be reproduced  as a special case of the general theory.  
At this stage,  we may no longer expect an essential role of canonical uncertainty relations in determining  a general form of amplification limits.

%
Another  topical aspect  on the probabilistic amplification  is its connection to   entanglement distillation.  On the one hand, the no-go theorem of Gaussian entanglement distillation tells us that  Gaussian operations 
 are unusable for  distillation of Gaussian entanglement
 \cite{Fiurasek,Giedke2002}. On the other hand, it has been shown that a specific design of non-deterministic linear amplifier (NLA) can enhance entanglement \cite{Ralph09}, and experimental demonstrations of entanglement distillation have been reported in  \cite{Chrzanowski2014,Xiang2010a}. Thereby, such an enhancement of entanglement could  signify a clear advantage of no-Gaussian operations over the Gaussian operations.   
Interestingly, a substantial difference between  an optimal amplification  fidelity  for deterministic  quantum gates  and that for probabilistic  physical processes has been  shown in Ref.~\cite{Chir13}. 
In there,   a standard Gaussian amplifier is identified as an optimal deterministic process for maximizing the fidelity, while  the NLA  turns out to   achieve the maximal fidelity for probabilistic gates in an asymptotical manner. 
However, these amplification fidelities  
 have not been associated with the context of entanglement distillation.
Hence, it is interesting  if one can find a legitimate amplification limit for Gaussian operations such that the physical process beyond the limit demonstrates the advantage of non-Gaussian operations.  More fundamentally, we may  ask whether  an amplification limit for Gaussian operations  could be derived as a consequence of the no-go theorem. %

The fidelity-based amplification limit \cite{Namiki11R,Chir13}  is defined on  
  an input-state ensemble called the \textit{Gaussian distributed coherent states}. This ensemble has been    utilized to demonstrate a non-classical performance of   continuous-variable (CV) quantum teleportation \cite{Furusawa98} and quantum memories \cite{julsgaard04a}. 
   The main idea underlying this ensemble is  to consider an effectively uniform set of input states in  a CV space  by using a Gaussian prior. 
We  can sample coherent states with modest input
power around the origin of the phase-space with a relatively flat prior while a rapid decay of the prior enables us to suppress the contribution of impractically high-energy input states.
  Given this ensemble, an experimental success criterion for CV gates is to surpass  the classical limit fidelity  due to  {\it entanglement breaking} (EB) maps \cite{Horo03a}. The classical fidelity was determined  for unit-gain channels in Ref.~\cite{Ham05} and for  lossy/amplification channels in Ref.~\cite{namiki07} (See also Ref.~\cite{Namiki11}). Further, the framework was  generalized to include whole  completely-positive (CP) maps, i.e., general quantum operations \cite{Chir13}.

Recently, a different form of such classical limits has been derived using an uncertainty product of canonical variables \cite{Namiki-Azuma13x}. It gives an optimal trade-off relation between canonical noises in order to outperform  EB maps for general amplification/attenuation tasks. 
This suggests that,  instead of the fidelity,  one can use  an uncertainty  product of canonical variables to evaluate the performance of amplifiers. 
 However,   for  a general amplification process,  it remains open   (i) how much excess  noise is  unavoidable  on canonical variables and (ii) whether  there exists a simple trade-off relation between noises of the canonical pair.

 In this paper  we resolve above questions by presenting an uncertainty-product form of amplification limits for general quantum operations based on the input ensemble of Gaussian distributed coherent states. It is directly derived by  using  canonical uncertainty relations and retrieves basic property of the traditional amplification limit. 
  We investigate attainability of our amplification limit and identify a parameter regime where Gaussian channels cannot achieve our bound but the NLA asymptotically achieves our bound. 
  We also point out the role of probabilistic amplifiers for entanglement distillation.  Using the no-go theorem for Gaussian entanglement distillation we find a condition that a probabilistic amplifier can be regarded as a local filtering operation to demonstrate entanglement distillation. This condition establishes a clear benchmark to verify an advantage of non-Gaussian operations beyond Gaussian operations with a feasible input set of coherent states and standard homodyne measurements.  

The rest of this paper is organized as follows. In section~\ref{OurLimit}, we present our amplification limit which is regarded as an extension of the traditional amplification limit \cite{amp}  for  two different  directions:    (i) It  determines the limitation with  an input ensemble of  a bounded power; (ii) 
It is applicable to stochastic quantum processes as well as quantum channels. 
  In section~\ref{AttainabilityAmplimit}, we consider attainability of our amplification limit  for Gaussian and non-Gaussian amplifiers.  In section~\ref{DistilBound}, we address the connection between our amplification limit and entanglement distillation.  We conclude this paper  with remarks in section~\ref{ConcRemark}.

\section{general amplification limits for Gaussian distributed coherent states} \label{OurLimit}
In this section we present a general amplification limit for Gaussian distributed coherent states which is  applicable to either probabilistic or deterministic quantum process. 
We review the fidelity-based  results of amplification limits in subsection~\ref{FiLimit} partly as an introduction of basic notations. We present our main theorem in subsection~\ref{DeLimit}. 

\subsection{Fidelity-based amplification limits}  \label{FiLimit}
We  consider  transmission of coherent states $\{\ket{\alpha }\}_{\alpha \in \mathbb{C}}$ 
 drawn from a Gaussian prior distribution  with an inverse width $\lambda >0, $ \begin{eqnarray}
p_\lambda( \alpha ) :=  \frac{\lambda }{\pi} \exp (- \lambda |\alpha |^2 ). \label{eq2}\end{eqnarray} 
We call the state ensemble $\{ p_\lambda (\alpha ) , \ket{\alpha}  \}_{\alpha \in { \mathbb{C}} }$ the Gaussian distributed coherent states.  A main motivation to use the Gaussian prior of Eq.~\eqref{eq2}  is to execute a uniform sampling of the input amplitude around the origin of the phase-space $| \alpha  |^2 \ll \lambda ^{-1} $ with keeping out the contribution of higher power input states for  $| \alpha  |^2  > \lambda ^{-1} $ by properly choosing the inverse width $\lambda >0$. A uniform average over the phase-space or an ensemble of completely unknown coherent states
 can be formally described by taking the limit $\lambda \to 0$. %

 Let us refer to   the following state transformation  as the phase-insensitive amplification/attenuation task of  a gain $\eta \ge0 $,  
\begin{eqnarray}
\ket{\alpha } \to  \ket{\sqrt \eta \alpha }. \label{task1}
 \end{eqnarray}  We say  the task is an amplification (attenuation)  if  $\eta \ge 1$ ($\eta <1$).  We may specifically call the task of $\eta =1$ the {\it unit gain} task.
We define an average fidelity of the phase-insensitive task for a physical map  $\mathcal E$  as 
\begin{eqnarray}
F_{\eta,\lambda} :=  {\int p_\lambda (\alpha) \bra{\sqrt \eta \alpha }   \mathcal E (\rho_{\alpha}) \ket{\sqrt \eta \alpha} d^2 \alpha}.  \label{fidelity}
\end{eqnarray}  Note that we use the following notation for the density operator of a coherent state throughout this paper: 
\begin{eqnarray} 
\rho _ \alpha = \ketbra{\alpha}{\alpha }. \label{abcs}
\end{eqnarray}
The fidelity-based amplification limit  \cite{Namiki11R, Chir13} is given as follows: For any quantum operation $\mathcal E $, i.e.,  a CP trace-non-increasing map,  it holds that  
\begin{eqnarray}
F_{\eta,\lambda}^{({ \rm Prob})}: = \frac{F_{\eta,\lambda}}{ P_s } \le 
\frac{1}{2}\left( \frac{1+ \lambda }{\eta}+ 1+  \left| \frac{1+ \lambda }{\eta}-1  \right|   \right) \label{Fresult1} ,
\end{eqnarray} 
where $P_s$ is  the probability that  $\mathcal E$ gives an output state  for the ensemble $\{p_\lambda (\alpha), \rho_\alpha \}$. It is defined as    
\begin{eqnarray}
P_s := { \tr\int p_\lambda (\alpha)     \mathcal E (\rho_{\alpha})  d^2 \alpha}. \label{defPS}
\end{eqnarray} 
As we will see  in the next subsection, this probability represents  a normalization factor when $\mathcal E$ acts on a subsystem of a two-mode squeezed state. 
Note that $P_s = 1 $  if $\mathcal E $ is a quantum channel, i.e.,  a CP trace-preserving map.  

In analogous to Eq.~\eqref{task1}, we may define a symmetric phase-conjugation task associated with the state transformation:   
\begin{eqnarray}
\ket{\alpha } \to \ket{\sqrt \eta \alpha^*  } .  \label{task2}
 \end{eqnarray}   Thereby, we may define an average fidelity of this task   as 
\begin{eqnarray}
F_{\eta,\lambda}^*  :=  {\int p_\lambda (\alpha) \bra{\sqrt \eta \alpha^* }   \mathcal E (\rho_{\alpha}) \ket{\sqrt \eta \alpha^* } d^2 \alpha}. \label{pfidelity}
\end{eqnarray} 
The fidelity-based phase-conjugation limit  is given by \cite{Namiki11R,Yang14} 
\begin{eqnarray}
{F_{\eta,\lambda}^* }^{({ \rm Prob})} =\frac{{F_{\eta,\lambda}^* } }{P_s}  \le \frac{1+ \lambda }{1+ \eta + \lambda }\label{Fresult2} 
,
\end{eqnarray} 

Note that one can generalize the fidelity-based quantum limits in Eqs.~(\ref{Fresult1})~and~\eqref{Fresult2}   for phase-sensitive cases by  introducing modified tasks  as
\begin{eqnarray}
\ket{\alpha } \to  S \ket{\sqrt \eta \alpha }, {\rm or  \ } \ket{\alpha } \to  S \ket{\sqrt \eta \alpha^* },
 \end{eqnarray} 
  where 
 \begin{eqnarray}
S:= S(r) = e^{r(\hat a^2- \hat a^{\dagger 2 })/2}\label{squeezeU} \end{eqnarray} 
  is a squeezing  unitary operation and $r$ represents the degree of squeezing. 
  The quantum limited fidelity values of Eqs.~\eqref{Fresult1}~and~\eqref{Fresult2} are  invariant under the addition of unitary operators since the optimal map can absorb the effect of  additional unitary operators  \cite{namiki07,Namiki12R,Namiki-Azuma13x}.

\subsection{Amplification limits via an uncertainty-product of canonical quadrature variables}  \label{DeLimit}

We may  consider a general phase-sensitive amplification/attenuation task  in terms of phase-space quadratures so that 
    average quadratures of the input coherent state $\rho_\alpha$ of Eq.~\eqref{abcs}  are transformed as 
\begin{align}
(x_\alpha, p_\alpha) \to (\sqrt{\eta_x}  x_\alpha, \sqrt{\eta_p}  p_\alpha),   \label{amppro} 
\end{align}
where   the gain pair of  the amplification/attenuation task $(\eta_x, \eta_p)$ is a pair of non-negative numbers,  and  the  mean quadratures for the coherent state $\rho_\alpha$ are defined as
\begin{eqnarray}
x_\alpha  := \tr (\hat x \rho_\alpha ) 
= \frac{\alpha + \alpha ^*}{\sqrt 2 }, \ p_\alpha := \tr (\hat p \rho_\alpha ) 
= \frac{\alpha - \alpha ^*}{\sqrt 2 i }.  \label{ShNote}  
\end{eqnarray}   
 Throughout this paper  we assume  the  canonical  commutation relation for canonical quadrature variables  $[\hat x, \hat p ]= i $,  which  is consistent with the standard relations such as 
 $\hat x = (\hat a + \hat a ^\dagger) / \sqrt 2 $, $\hat p =  (\hat a - \hat a ^\dagger) / (i \sqrt 2 )$, and     $\hat a \ket{\alpha} = \alpha \ket{\alpha} $.  
Similarly to Eq.~\eqref{amppro}, we may consider  a general  phase-conjugation task associated with the following transformation: 
\begin{align}
(x_\alpha, p_\alpha) \to (\sqrt{\eta_x}  x_\alpha, - \sqrt{\eta_p}  p_\alpha).  \label{pstask}
\end{align}

Given the task of  Eq.~\eqref{amppro}, we may measure the performance of an amplifier $\mathcal E$ by using the square deviation, 
\begin{align}
  \tr [  (\hat z  - \sqrt \eta_ z  z_\alpha)^2  \mathcal E (\rho_\alpha)],  \label{dev}
\end{align}  where  $z \in \{x, p\}$.  Note that, if the mean output quadratures are equal to the output of the transformation of Eq.~\eqref{amppro}  as $ \tr [ \hat z    \mathcal E (\rho_\alpha )]= \sqrt \eta_z z_ \alpha   $, the expression of Eq.~\eqref{dev}  turns to the variance of the output quadrature
\begin{align}
  \tr [  (\hat z  - \sqrt \eta_ z  z_\alpha)^2  \mathcal E (\rho_\alpha)] &=   \tr [  \hat z ^2    \mathcal E (\rho_\alpha)]  -  (\tr [   \hat z  \mathcal E (\rho_\alpha)]) ^2 \nonumber \\  &= \ave{\Delta ^2 \hat z }_{\mathcal E (\rho_ \alpha )}  .  
\end{align}  However, it is impractical  to consider that   the linearity of the  transformation 
\begin{align}
\tr [ \hat z    \mathcal E (\rho_\alpha )]= \sqrt \eta_z z_ \alpha  \label{liniconcon} \end{align}  holds in experiments for every input amplitude $\alpha \in \mathbb{C}$.  We thus proceed our formulation  {\it without} using this condition.

Instead of the point-wise constraint on $\alpha$, we consider an average of the quadrature deviations  with the  Gaussian prior distribution $p_\lambda $ of Eq.~\eqref{eq2}.   We seek for the physical process that minimizes   the {\it mean square deviations} (MSD) of canonical quadratures: \cite{Namiki-Azuma13x} 
\begin{eqnarray}
 \bar V_x (\eta, \lambda) &:=& \tr \int  p_\lambda (\alpha )   (\hat x-  \sqrt \eta x_\alpha )^2   \mathcal E ( \rho _\alpha ) d^2 \alpha , \nonumber \\ 
  \bar V_p (\eta, \lambda) &:=  &  \tr \int  p_\lambda (\alpha )   (\hat p \mp   \sqrt \eta p_\alpha )^2   \mathcal E ( \rho _\alpha ) d^2 \alpha,   \label{defbarxp} \end{eqnarray}
  where the lower sign of the second expression is  for the case of the phase-conjugation task in Eq.~\eqref{pstask}. 
The MSDs of  Eq.~\eqref{defbarxp} can be observed experimentally by measuring the first and the second moments of the quadratures $\{\hat x,  \hat  p, \hat x^2, \hat p^2 \}$ for the output of the physical process $\mathcal E (\rho_ \alpha)$.  Due to canonical uncertainty relations,  $\bar V_x$ and  $\bar V_p$  could not be arbitrary small, simultaneously. We can find a rigorous trade-off relation between $\bar V_x$ and  $\bar V_p$  from the following theorem. %

\textbf{Theorem 1.--- }
For any given $\eta_x > 0 $,  $\eta_p > 0 $, and $\lambda >0 $, any quantum operation (or stochastic quantum channel) $\mathcal E$ satisfies  
\begin{eqnarray}\prod_{z = x, p } \left[ \frac{ \bar V_z (\eta_z, \lambda) }{P_s}- \frac{\eta_z}{2 (1+ \lambda )} \right] 
 \ge \frac{1}{4}\left|   \frac{ \sqrt{\eta_x \eta_p}  }{1+ \lambda } \mp 1  \right|^2 \label{AUP2} \end{eqnarray} where  $  P_s$   and $\bar V_z $ are defined in 
Eqs.~\eqref{defPS}~and~\eqref{defbarxp}, respectively.  
Moreover, the lower signs of Eqs.~\eqref{defbarxp}~and~\eqref{AUP2} correspond to the case of  the phase-conjugation task in Eq.~\eqref{pstask}.  

\textbf{Proof.}---Let $J= J_{AB}$ be a density operator of a two-mode system $AB$ described by  $ [\hat x_A, \hat p_A ] =  [\hat x_B, \hat p_B ]  =i   $. 
  Canonical  uncertainty relations and property of variances lead to
\begin{eqnarray}
 &&\tr [ (\hat x_A-  g_x \hat x_B )^2 J ] \tr [ (\hat p_A+   g_p \hat p_B )^2 J ]  \nonumber \\ 
  & \ge  & \ave{\Delta ^2 (\hat x_A-  g_x \hat x_B )}_J \ave{\Delta ^2 (\hat p_A+   g_p \hat p_B )}_J  \ge \frac{1}{4}(1-g_x g _p)^2.   \nonumber \\ \label{usr1}\end{eqnarray}
  Here, we will prove the case of the normal amplification/attenuation process by assuming $g_x  \ge 0$ and   $g_p \ge 0$. The proof for the phase-conjugation process runs  similarly by considering the  case of $g_x \ge 0 $ and $ g_p <0$.
  
From a standard notation  
$\hat x_B =   {(\hat b + \hat b^\dagger )}/{\sqrt 2 }$ 
and  the cyclic property of the trace we can write
\begin{eqnarray}
 &&\tr [ (\hat x_A-  g_x \hat x_B )^2 J ] \nonumber \\ 
  &=  & \tr [  \hat x_A^2 { J}  - 2 g _x \hat x_A \left(\frac{\hat  b ^\dagger  { J} +  {J} \hat  b }{\sqrt 2 }\right) \nonumber \\
  &&+ \frac{{g_x}^2}{2} (\hat b ^{\dagger 2 }  {  J} +  { J} \hat b ^2  +2 \hat b ^{\dagger  }  {  J} \hat b - J  ) ]  \nonumber \\
 &=& \tr_A \int\frac{d^2 \alpha }{\pi}(\hat x_A -  g_x x_\alpha )^2 \bra{\alpha ^*}J \ket{\alpha ^*}_B  -\frac{g_x^2}{2}, \label{ForX}\end{eqnarray}
where, in the final line, 
we execute the partial trace  by $\tr_B [\ \cdot \ ] \to \int   \bra{\alpha ^ *} \cdot   \ket{\alpha ^ *}_B  \frac{d^2 \alpha  }{\pi} $ and  use  the property of the coherent state
   $ \hat b \ket{\alpha ^ *}_B = \alpha ^* \ket{\alpha ^ *}_B $  and $  \bra{\alpha ^ *}_B \hat b ^\dagger  = \alpha \bra{\alpha ^ *}_B $.
Similarly, starting from  $\hat p_B =i {(\hat b^\dagger -  \hat b)}/{\sqrt 2 }$ we have 
\begin{eqnarray}
 &&\tr [ (\hat p_A +  g_p \hat p_B )^2 J ] \nonumber \\ 
  &=& \tr_A \int\frac{d^2 \alpha }{\pi}(\hat p_A -  g_p p_\alpha )^2 \bra{\alpha ^*}J \ket{\alpha ^*}_B  -\frac{g_p^2}{2}.  \label{ForP}\end{eqnarray}
    Next, suppose that $J$ is prepared by an action of a quantum operation 
    $\mathcal E$  as $J =  \mathcal E_A  \otimes  I_B  ( \ketbra{\psi_\xi}{\psi_\xi} ) /P_s $ where  $ \ket{\psi_\xi}= \sqrt{1-\xi^2 }\sum_{n=0}^\infty \xi ^n  \ket{ n} \ket{ n }$ is a two-mode squeezed state with $\xi \in (0,1)$ and $P_s:=  \tr[\mathcal E_A  \otimes  I_B  ( \ketbra{\psi_\xi}{\psi_\xi} )]$.
This implies $\bra{\alpha ^*}J \ket{\alpha ^* }_B  = (1-\xi^2) e ^{-(1-\xi^2)|\alpha |^2 }\mathcal E_A( \rho_{ \xi\alpha}  )/ P_s$.
From this relation and Eqs.~(\ref{eq2}),~(\ref{ForX}),~and~(\ref{ForP}) we obtain 
\begin{eqnarray}&&
 \tr [ (\hat p_A +  g_p \hat p_B )^2 J ]
\tr [ (\hat x_A-  g_x \hat x_B )^2 J ] \nonumber \\
 &=& \prod_{z=x,p} \left(  \tr_A \int\frac{d^2 \alpha }{\pi}(\hat z_A -  g_z z_\alpha )^2 \bra{\alpha ^*}J \ket{\alpha ^*}_B  -\frac{g_z^2}{2} \right) \nonumber \\
   &=& \prod_{z=x,p}  \left[\frac{\tr  \int p_\lambda (\alpha )  ( \hat z   -  \sqrt \eta_z z_ {\alpha}   )^2 \mathcal E (\rho_{ \alpha }){d^2 \alpha }}{\tr \int p_\lambda (\alpha )  \mathcal E (\rho_{ \alpha }){d^2 \alpha }}  -\frac{\eta_z}{2(1+ \lambda) }\right],  \nonumber \\ \label{LHS} \end{eqnarray}
where, in the final step, we drop the subscript $A$, rescale the integration variable as $\xi \alpha \to  \alpha $, and introduce 
\begin{eqnarray}
\eta_x = (1+ \lambda )g_x^2 ,\ \eta_p = (1+ \lambda )g_p^2 ,\ \lambda = \frac{1- \xi ^2 }{ \xi ^2 }. \label{paragx}
\end{eqnarray}
Finally, concatenating Eqs.~(\ref{defbarxp}), (\ref{usr1}), (\ref{LHS}), and (\ref{paragx}), we can reach our theorem 1 of Eq. (\ref{AUP2}). 
  \hfill$\blacksquare$ 

   \begin{figure*}[phbt]
    \includegraphics[width=0.4\linewidth]{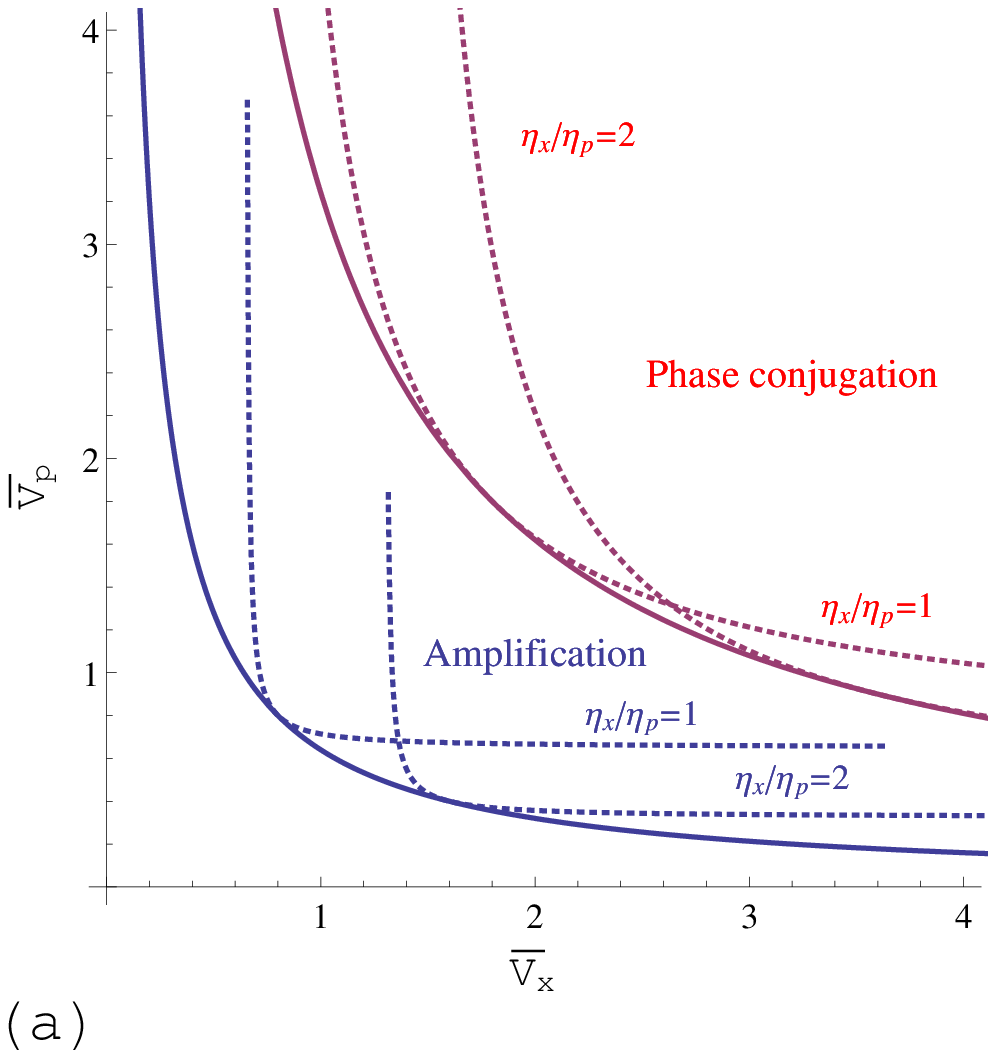}
    \includegraphics[width=0.4\linewidth]{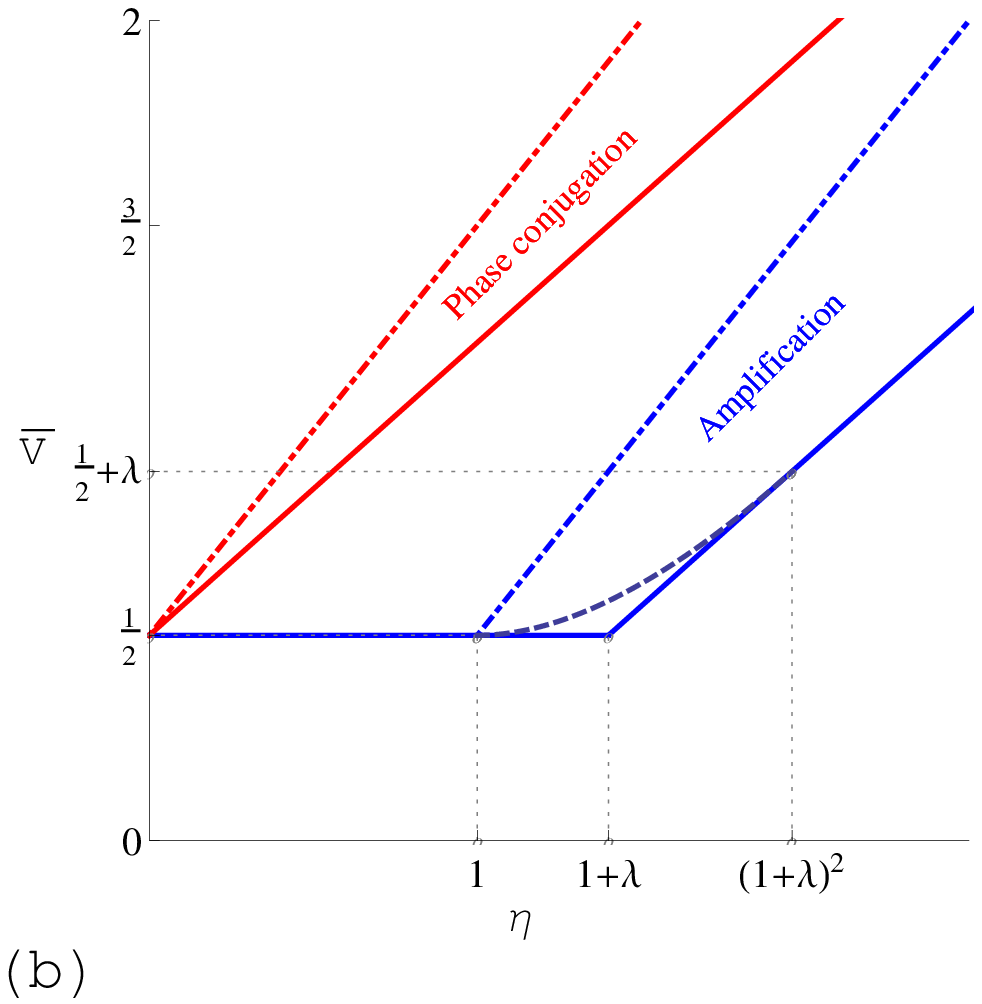}
  \caption{(Color Online) 
  (a)  
 Solid curves represent  general lower bounds on the  product of  the mean square deviations (MSD)  of Eq.~\eqref{defbarxp} given by $\bar V _x \bar V _p \ge (\eta^\prime + |\eta^\prime \mp 1|)^2/4$   [See Eq.~\eqref{combo}]. Here, an effective gain factor  is set to be $ \eta ^\prime  = {\eta}/({1 +\lambda})  = 1.3$ with the constraint on the gain pair $\sqrt{ \eta_x \eta_p } =\eta $. This constraint parameterizes the gain pair as $(\eta_x, \eta_p) = (\eta e^R,\eta e^{-R})$ with $R \in \mathbb{R}$.   
  The upper solid curve is for the phase-conjugation task and the lower solid  curve is for the normal amplification task. Each of them can be determined by taking the intersection of our amplification limits  in Eq.~\eqref{AUP2} with various ratios between the gain pair $\eta_x / \eta_p =e^{2R}$. Dotted curves represent the  cases for  $\eta_x / \eta_p =1$ and  $\eta_x / \eta_p =2$. 
   (b) Typical behavior of the MSDs in the cases of $\bar V_x = \bar V_p = \bar V$ as a function of the gain $\eta$.      The upper solid line represents the limitation on the phase-conjugation process [Eq.~\eqref{eeeto}] and the lower kinked solid line represents the limitation on the  normal amplification/attenuation process [Eq.~\eqref{eeto}].   The dashed curve for $\eta \in [1, (1+\eta)^2 ]$ shows the minimum of the MSD  due to Gaussian channels $\bar V ^\star$  [Eq.~\eqref{result1}].    Dash-dot lines are due to the traditional form of amplification limits for completely unknown coherent states [Eqs.~(\ref{pil})~and~(\ref{pil2}) with $G = \eta$].  They can be retrieved by our bounds in the limit of $\lambda \to 0$. In this figure we set $\lambda = 0.4$ so that the structure around $\eta = 1 +\lambda $ is  displayed clearly.   When $\eta =0$, all lines  indicate the minimum value of $\bar V = 1/2$ due to  canonical uncertainty relations as it corresponds to the trivial case of $g_x =g_p = 0 $ in Eq.~\eqref{usr1}. }
    \label{fig:ampfig1.eps}
\end{figure*}

Our theorem 1 states that any physical map is unable to break  the uncertainty-relation-type  trade-off inequality for quadrature deviations on average.   
It draws an inverse-proportional curve in the $\bar V _x $-$\bar V _p$~plane with a given  pair of  $(\eta_x,\eta_p)$, and the area below the curve is unattainable by any quantum process including probabilistic amplifiers [See FIG.~\ref{fig:ampfig1.eps}(a)].  Equation~\eqref{AUP2}  is essentially the same structure as the traditional form of amplification limits \cite{amp} [see. Eq.~\eqref{AUR0} of Appendix~\ref{AP1}].  However,  note that our theorem can be applied to  probabilistic amplifiers. In addition, it holds without the linearity condition of Eq.~\eqref{liniconcon}. Nevertheless,   it retrieves the traditional expression in the limit of $\lambda \to 0$.   
A detailed interrelation between our theorem 1 and the traditional amplification limit can be found in Appendix~\ref{AP1}.

In order to see the role of our amplification limit for the case of   the phase-sensitive process, we may consider the curve in the $\bar V _x $-$\bar V _p$~plane  with a different set of $(\eta_x, \eta_p)$ under  the constraint of a fixed gain $\eta = (\eta_x \eta_p ) ^{1/2}$  as in FIG.~\ref{fig:ampfig1.eps}(a). Then, the intersection of the unattainable area can be represented by another    inverse-proportional curve. 
 This curve determines the minimum uncertainty in the $\bar V _x $-$\bar V _p$~plane similar to 
   %
   the minimum uncertainty curve for  normal squeezed coherent states. 
  In fact, we can show an expression that the minimum of the product $\bar V _x \bar V_p$ is bounded from below by  a constant as follows.      
 Let us parameterize   the boundary of Eq.~\eqref{AUP2}  
 as   
 \begin{align} 
(\bar V_x ,  \bar V_p  )  = \frac{1}{2}\left|  \frac{\sqrt{ \eta_x \eta_p} }{1+ \lambda } \mp 1 \right|  ( e^R, e^{-R}) + \frac{1}{2(1+ \lambda )}(\eta_x , \eta_p ), \label{parame}  
\end{align} where $R \in \mathbb{R}  $. Suppose that the gain is fixed as $\eta = \sqrt{\eta_x \eta_p}$. Then, we can write $\eta_x = \eta e^{-2r}$ and $\eta_p = \eta e^{ 2r}$ with $r \in \mathbb{R}$. Hence, we have
 \begin{align} 
 \bar V_x   \bar V_p  = & \frac{1}{4}\left\{ {{\eta^\prime }^2 }  +   \left|  {\eta^\prime }   \mp 1 \right| ^2 +  \left| {\eta^\prime } \mp 1 \right|  \eta ^\prime   ( e^{R+2r }  +  e^{- R-2r } ) \right\}
\nonumber\\ &\ge \frac{1}{4} \left( \eta^\prime +  \left| {\eta^\prime }  \mp 1 \right|  \right)^2 , \label{combo} 
\end{align}   where we defined $\eta ^\prime = \eta /(1+ \lambda )$ and used $ e^{R+2r }  +  e^{- R-2r }  \ge 2$.  This gives the lower bound of the  uncertainty product   $\bar V_x   \bar V_p  $ under the constraint of the fixed gain $\eta = \sqrt{\eta_x \eta_p}$, and it implies an inverse-proportional relation between $\bar V_x$ and $\bar V_p$ shown in FIG.~\ref{fig:ampfig1.eps}(a). 
Note that the boundary of Eq.~\eqref{combo} is  parameterized as 
 \begin{align} 
(\bar V_x ,  \bar V_p  )  = \frac{1}{2} \left( \eta^\prime +  \left| {\eta^\prime }^2 \mp 1 \right|  \right)   ( e^R, e^{-R}) .  \label{parapara}
\end{align} This expression is obtained by substituting $\eta_x= \eta  e^R$ and $\eta_p= \eta  e^{-R}$ into Eq.~\eqref{parame}. 
We will discuss the design of physical amplifiers that potentially  achieve this boundary in the next section.

It would be instructive to illustrate the gain dependence of our quantum limit for simple cases [See  FIG.~\ref{fig:ampfig1.eps}(b)]. 
For the symmetric  case  with $\eta_x= \eta_p = \eta $ and  $\bar V_x = \bar V_p = \bar V $, we can write our theorem~1 of Eq.~\eqref{AUP2} for the normal amplification/attenuation task as  
\begin{eqnarray}
  \frac{\bar V}{P_s}  \ge   \frac{1}{2} \left( \frac{\eta}{1+ \lambda }+  \left | \frac{\eta}{1+ \lambda } -1 \right | \right) \label{eeto}, 
 \end{eqnarray} 
or equivalently,
\begin{eqnarray} \frac{\bar V}{P_s}  \ge    \begin{cases}\frac{1}{2} &   \eta  \in [0, 1+ \lambda ]   \\  \frac{\eta }{  1 + \lambda  }  -\frac{1}{2}   &   \eta   >  1+ \lambda     \end{cases}  .  \label{Fresult000}  
\end{eqnarray}  This  shows  basically  the same structure in 
the expression of the fidelity-based amplification limit in   Eq.~\eqref{Fresult1}.  For phase-conjugation, we alternately have 
\begin{eqnarray}
\frac{\bar V}{P_s}  \ge   
   \frac{\eta }{  1 + \lambda  }+\frac{1}{2} . \label{eeeto}
 \end{eqnarray}  
The minima of the MSDs $\bar V$  for both of Eqs.~\eqref{eeto}~and~\eqref{eeeto} are shown as functions of $\eta$ in FIG.~\ref{fig:ampfig1.eps}(b). They obviously fall below the lines due to the traditional form of amplification limits given in Eqs.~\eqref{pil}~and~\eqref{pil2} of Appendix~\ref{AP1}  for $ \eta > 1$ in the case of the normal amplification task and for 
 $\eta > 0$ in the case of the phase-conjugation task, respectively. 
Note that the gap disappears in the limit of $\lambda \to  0$ although it is impossible to test amplification devices for completely unknown coherent states in the real world.


As we have already mentioned, the MSDs of  Eq.~\eqref{defbarxp} can be observed experimentally by measuring the first and the second moments of the quadratures $\{\hat x,  \hat  p, \hat x^2, \hat p^2 \}$ for the output of the physical process $\mathcal E (\rho_ \alpha)$.  This can be done by standard homodyne measurements. In contrast, one need to know higher order moments of the quadratures in order to determine the fidelity to coherent states in Eqs.~\eqref{fidelity}~and~\eqref{pfidelity}   when homodyne measurements are performed. This is because the output state $\mathcal E (\rho_ \alpha)$ could be a non-Gaussian state. 
Note that one can find a lower bound of the fidelity from the observed value of the MSDs   \cite{namiki07,Namiki-Azuma13x}.

\section{Achievability of the  amplification limit} \label{AttainabilityAmplimit}
In this section, we consider attainability of our amplification limit given in Eq.~\eqref{AUP2} by using a standard Gaussian amplifier and a probabilistic amplifier. 

\subsection{Gaussian amplifier}
In this subsection we investigate  the performance of Gaussian channels for the  normal  amplification/attenuation process (See subsection~\ref{ConjugateAmplification} for the phase-conjugation process). 

At a moment, let us consider the phase-insensitive case, i.e., 
 $\eta_x=\eta_p =\eta >0$.
The quantum limited phase-insensitive Gaussian amplifier/attenuator with the gain $G$ 
transforms the first and second moments of quadratures \cite{Hol08} as
\begin{eqnarray}\tr{[\hat z \mathcal A_G (\rho_\alpha )]}&=& \sqrt { G} z_\alpha \nonumber , \\ \tr{[\hat z ^2 \mathcal A_G (\rho_\alpha )]}&=&  G  z_\alpha^2 +(G+ |G-1 |)/2,  \end{eqnarray} where $z \in \{x, p\}$ and we use the notation in Eqs.~\eqref{abcs}~and~\eqref{ShNote}. This yields the following expression for the MSDs of Eq.~\eqref{defbarxp},
\begin{eqnarray}
\bar V_z (\eta,\lambda)\Big |_{\mathcal E = \mathcal A_G} =  \frac{1}{ \lambda}  (\sqrt{  G} - \sqrt \eta  ) ^2 +\frac{G+|G-1 |}{2}. \label{eq12} \end{eqnarray} 
When the prior distribution $p_\lambda (\alpha )$ of  Eq.~\eqref{eq2} becomes broader so that $\lambda \to 0 $, the contribution of the first term of Eq.~\eqref{eq12} becomes significantly larger. In this limit, $G =\eta $ is the solution that minimizes the MSDs and the optimality of the Gaussian amplifier is retrieved, namely, the Gaussian amplifier $ \mathcal A_G $ saturates our bound of Eq.~\eqref{AUP2}  similar to that   $ \mathcal A_G $ saturates the traditional amplification limit in  Eq.~(\ref{pil}). 

 In order to minimize the MSDs for a finite distribution with $ \lambda >0 $ we may rewrite Eq.~\eqref{eq12} as 
\begin{eqnarray}
\bar V_z  (\eta,\lambda) =  & & 
 \begin{cases}
   \frac{1 }{ \lambda} \left(\sqrt{  G} - {\sqrt \eta } \right) ^2 + \frac{1}{2} &   G \in [0, 1] , \\ 
  \frac{1+ \lambda}{ \lambda} \left(\sqrt{  G} -\frac{\sqrt \eta }{1+ \lambda} \right) ^2 + \frac{\eta  }{1+ \lambda} - \frac{1}{2}  & G > 1.  \end{cases}   \nonumber 
\end{eqnarray} 
For the first case of $G \in [0, 1] $, $G =\eta$ fulfills the equality of  Eq.~(\ref{AUP2}) for $\eta  \in [0, 1]$. 
For the second case of $G > 1  $, the optimal gain $ G = \eta /(1+\lambda )^2  $  fulfills the equality of  Eq.~(\ref{AUP2}) for $\eta  \ge (1+ \lambda )^2 $. 
Thereby, the minimum MSD due to Gaussian channels is divided into the following three cases [See   FIG.~\ref{fig:ampfig1.eps}(b)], \begin{eqnarray}
 \bar V_z ^\star  (\eta,\lambda) =   
 \begin{cases}
 \frac{1}{2} &   \eta  \in [0, 1] , \\ 
  \frac{1 }{ \lambda} \left( {\sqrt \eta } -1 \right) ^2 + \frac{1}{2} &   \eta \in \left( 1 , (1+ \lambda)^2\right) , \\    \frac{\eta  }{1+ \lambda} - \frac{1}{2}  & \eta  \ge (1+ \lambda)^2.  \end{cases}   &  \label{result1} 
\end{eqnarray} 
Hence, in the phase-insensitive  case of the normal  amplification/attenuation process, we can conclude that the Gaussian channel constitutes an optimal quantum device that saturates our amplification limit except for  the range of the gain factor  $ \eta \in \left( 1 , (1+ \lambda)^2\right)$.

To proceed the case of an asymmetric pair of gains,     we can choose $\eta_p > \eta_x >0 $ without loss of generality.  Let us write 
 $\eta = \sqrt{\eta_x \eta _p}  $  with  \begin{eqnarray}
\eta_x=  \eta e^{-2r}, \ \eta_p=  \eta e^{2r}, \ r  > 0 . \label{asgain}
\end{eqnarray}
We can readily see that an action of the quadrature squeezer $S$ of Eq.~\eqref{squeezeU}  followed by the amplification process modifies the first and second moments as 
\begin{eqnarray}
\tr{[\hat x S\mathcal A_G (\rho_\alpha )S^\dagger]}&=& e^{-r}\tr{[\hat x \mathcal A_G (\rho_\alpha )]}, \nonumber \\
 \tr{[\hat p S\mathcal A_G (\rho_\alpha )S^\dagger]}&=&  e^{r}\tr{[\hat p \mathcal A_G (\rho_\alpha )]}, \nonumber \\ 
\tr{[\hat x ^2 S \mathcal A_G (\rho_\alpha ) S^\dagger]}&=&   e^{-2r}\tr{[\hat x^2 \mathcal A_G (\rho_\alpha )]}, \nonumber \\ 
\tr{[\hat p ^2 S \mathcal A_G (\rho_\alpha ) S^\dagger]}&=&   e^{2r}\tr{[\hat p^2 \mathcal A_G (\rho_\alpha )]}. \label{acase}
\end{eqnarray} 
From Eqs.~(\ref{asgain})~and~(\ref{acase}), we can observe that the Gaussian channel $\mathcal E (\rho) = S \mathcal A_G (\rho) S^\dagger$  fulfills
\begin{align}
\bar V_x(\eta_x,\lambda ) -\eta_x/2 &=e^{- 2r }[\bar V_x(\eta,\lambda ) -\eta/2 ], \nonumber \\ 
\bar V_p(\eta_p,\lambda ) -\eta_p/2 &=e^{2r }[\bar V_p(\eta,\lambda ) -\eta/2 ]. \label{36}\end{align}  This relation with the expression of Eq.~(\ref{result1}) implies that the channel $\mathcal E (\rho) = S \mathcal A_G (\rho) S^\dagger$   saturates our  quantum limit  Eq.~(\ref{AUP2}) 
 except for $ \eta \in \left( 1, (1+\lambda)^2 \right)$.

  Consequently, Gaussian channels constitute optimal physical processes in the amplification/attenuation task under  the practical setting of the Gaussian distributed coherent states 
   unless the normalized gain factor  is in the proximity of $\eta/ (1+\lambda) \sim 1$. In this sense, we could keep the term of the ``quantum-limited process''  or ``quantum limit amplifier'' for the Gaussian amplifier  $\mathcal A_G$. Similar statements hold for fidelity-based results \cite{Namiki11R,Chir13}.  
Note that our analysis here does not preclude the possibility that a trace-decreasing Gaussian amplifier could achieve the bound for $ \eta \in \left( 1, (1+\lambda)^2 \right)$, although it seems unlikely that the trace-decreasing class has an advantage as we will discuss later in section \ref{DistilBound}.

 \subsection{Non-Gaussian amplification}      \label{NGamp}
In this subsection we investigate the performance of a non-Gaussian  operation, the NLA  of Ref.~\cite{Ralph09}, for  the normal amplification process. 
We will show that the performance of the NLA approaches  arbitrarily close to our amplification limit of Eq.~\eqref{AUP2}  for the range of the gain   $\eta \in (1, (1+ \lambda)^2 )$,  %
  where 
    the Gaussian amplifier shows a substantially lower performance 
as in 
FIG.~\ref{fig:ampfig1.eps}(b).

Let us consider the probabilistic amplifier described by $\mathcal Q_g (\rho ) \propto   Q_N \rho Q_N$ with  
$ Q_N  = {\cal N }^{1/2} \sum_{n=0}^N g^n \ketbra{n}{n}$ where  we assume $g \ge 1 $ and ${\cal N} >  0 $.  This  leads to
\begin{eqnarray}
 Q_N \ket{\alpha} &=&  {\cal N }^{1/2} e^{-|\alpha|^2 /2 } \sum_{n=0}^N \frac{(g\alpha )^n}{\sqrt{n!}}\ket{n}=: \ket{\omega_{g,N,\alpha} }, \nonumber \\
\hat a \ket{\omega} &=&g\alpha e^{-|\alpha|^2 /2 } \sum_{n=0}^{N-1} \frac{(g\alpha )^n}{\sqrt{n!}}\ket{n}. \label{alphaW} \end{eqnarray}Hence,  we can write $Q_N \ket{\alpha} \propto \ket{g \alpha}$ on the truncated photon-number space $\{\ketbra{n}{n} \}_{n= 0,1,2, \cdots , N }$, and the operation $\mathcal Q_g$ amplifies coherent states without extra noises in the limit $N \to \infty$.   
The trace-non-increasing condition for quantum operations $  Q_N ^2 \le  \openone$  implies ${\cal N} \le  g^{-2 N }$.   
In what follows we focus on the phase-insensitive case of $\eta_x = \eta_p = \eta$.  The case of the phase-sensitive process with a possibly asymmetric gain pair $(\eta_x,\eta_p)$ can be addressed  by repeating the procedure of the previous subsection.

 From Eq. (\ref{alphaW}) we can easily calculate the mean values  $\ave{\hat x }_\omega$, $\ave{\hat p }_\omega$,  and $\ave{\hat a^\dagger \hat a   }_\omega = \ave{\hat x^2 +\hat p^2   -1 }_\omega /2 $.  
 As a consequence we  can obtain the following expression: 
\begin{eqnarray}
&&\ave{(\hat x - \sqrt \eta x_\alpha )^2 +(\hat p - \sqrt \eta p_\alpha )^2}_\omega \nonumber \\
&=&\ave{\hat x^2 + \hat p^2}_\omega -2 \sqrt\eta \ave{\hat x x _\alpha +\hat p p _\alpha }_\omega + \eta (x_\alpha ^2  + p_\alpha ^2) \braket{\omega}{\omega}\nonumber \\
&=&  { \cal N  }   \Bigg[ 2(g^2 -2 \sqrt \eta g)|\alpha |^2 \sum_{n=0}^{N-1} \frac{(g^2 |\alpha| ^2)^{n}}{n!} \nonumber \\
&& \  +  \left(2 \eta |\alpha |^2 +1 \right) \sum_{n=0}^{N} \frac{(g^2 |\alpha| ^2)^{n}}{n!} 
  \Bigg] e^{-|\alpha |^2}  \label{x-psq},  
\end{eqnarray} where $( x_\alpha , p_\alpha)$ is given by Eq.~\eqref{ShNote}. 

Now, let us evaluate the mean square deviations (MSD) of Eq.~(\ref{defbarxp}) for the probabilistic amplifier $\mathcal Q_g$ [The physical process is given by  $\mathcal E (\rho)= \mathcal Q_g  (\rho ) $]. Due to its  phase-insensitivity, we can write 
$\bar V: =(\bar V_x+ \bar V_p)/2= \bar V_x =\bar V_p$.
 Using this relation and Eq.~\eqref{x-psq} we have  
\begin{eqnarray}
\bar V&=&\frac{1}{2}\int p_\lambda(\alpha)\ave{(\hat x - \sqrt \eta x_\alpha )^2 +(\hat p - \sqrt \eta p_\alpha )^2}_\omega  d^2\alpha   \nonumber \\
&=&  \frac{ \cal N \lambda}{1+\lambda}   \Bigg[ 
(g - \sqrt \eta)^2 \sum_{n=0}^{N-1} \frac{g^{2n} (n+1)}{(1+\lambda )^{n+1}} +   \eta \cdot  \frac{g^{2N} (N+1)}{(1+\lambda )^{N+1}} \nonumber \\
&& ~ ~ ~ ~~ ~ + \frac{1}{2} \sum_{n=0}^{N} \frac{g^{2n} }{(1+\lambda )^n} \label{ngmsd}
  \Bigg] .
\end{eqnarray}
In this expression and the following expression, the  
  integrations can be  calculated by using   
$\int p_\lambda(\alpha) e^{-|\alpha |^2}  |\alpha |^{2k} d^2\alpha =  \lambda {k!}/{(1+ \lambda )^{k+1}}$ with $p_\lambda$ of Eq.~\eqref{eq2}. 
We can write the probability  that the NLA operation  ${\cal Q}_g $ gives an output in  Eq.~\eqref{defPS} as    
\begin{align}
P_s &= \tr   \int p_\lambda (\alpha) Q_N \ketbra{\alpha}{\alpha} Q_N d^2\alpha    \nonumber \\ 
&=  \frac{ \cal N \lambda}{1+\lambda}    \sum_{n=0}^{N} \frac{g^{2n} }{(1+\lambda )^n} \ge \frac{ \cal N \lambda}{1+\lambda}    . \label{eq24}
\end{align}
 As we have seen in section~\ref{OurLimit}, this probability $P_s$ corresponds to the physical probability that the amplifier gives the desired outcome when it acts on a subsystem of a two-mode squeezed state. 

Our concern here is the parameter regime of the gain factor $\eta  \in (1, (1+ \lambda)^2 )$ where the Gaussian channel cannot achieve our quantum limitation of Eq.~\eqref{AUP2} [See FIG.~\ref{fig:ampfig1.eps}(b)]. We will address this regime by further dividing it into two sub-regimes $\eta  \in (1, 1+ \lambda )$ and $\eta  \in (1+ \lambda, (1+ \lambda)^2 )$ since the behavior of the minimum MSD suddenly changes at $\eta = 1+ \lambda $.

For $\eta  \in (1, 1+ \lambda )$, by substituting  $g = \sqrt \eta $ into  Eqs.~(\ref{ngmsd})~and~\eqref{eq24} we obtain 
the form of the MSDs for the probabilistic amplifier  ${\cal Q}_g $ as  \begin{eqnarray}
\bar V^{(\textrm{Prob})}&= & \frac{\bar V}{P_s} \nonumber \\ 
&=&  
\frac{ {\cal N } \lambda  \eta   }{P_s} \left( \frac{\eta }{ 1+\lambda  }\right)^{N}\frac{  (N+1)}{(1+\lambda )^2}   +  \frac{1}{2}. \end{eqnarray}
Since $\eta /(1+ \lambda ) <1$ and $P_s$ is  bounded from below as in Eq.~\eqref{eq24}, we have $ \bar V^{(\textrm{Prob})} 
= 1/2 $ for $ N \to \infty$. This concludes  that the NLA $\mathcal Q_g$  saturates our bound of Eq.~\eqref{Fresult000}  for $\eta \in (1, 1+ \lambda ) $.

For $\eta  \in \left( 1+\lambda , (1+ \lambda )^2\right)$, let be $g =(1+ \lambda)/\sqrt \eta \ge 1 $ and $x:= (1+ \lambda )/ \eta < 1 $. Then, we can respectively rewrite Eqs.~(\ref{ngmsd})~and~\eqref{eq24}  as \begin{align}\bar V =& \frac{\cal N \lambda }{1+ \lambda }  \left(  \frac{1}{x} + Nx^N + \frac{1}{2} \sum_{n=0}^{N} x^n \right),\nonumber \\  P_s=& \frac{\cal N \lambda }{1+ \lambda }  \sum_{n=0}^{N} x^n.  \end{align}   From these expressions we obtain 
\begin{eqnarray}
\bar V^{(\textrm{Prob})}=  \frac{\bar V}{P_s} 
&=&  \underbrace{ \frac{1}{ 1 - x^{N+1}}}_{\ge 1 } \left(\frac{1}{x} -1+ \underbrace{N (1-x)x^N}_{\ge 0 }\right)+\frac{1}{2} \nonumber .  \end{eqnarray}
From this expression and $x <1$, we obtain 
\begin{eqnarray}
\lim_{N \to \infty} \frac{\bar V}{P_s} = \left(\frac{1}{x} -1\right)+\frac{1}{2} =  \frac{\eta}{1+ \lambda } -\frac{1}{2}.
 \end{eqnarray}
This coincides with   our bound  in Eq.~(\ref{Fresult000}) when $\eta  \in \left( 1+\lambda , (1+ \lambda )^2\right)$. Therefore, 
 one can design the probabilistic machine whose performance 
is arbitrary close to the amplification  limit  of Eq.~\eqref{AUP2}  by taking sufficiently large $N$, both in the sub-regimes $\eta \in (1,  1+\lambda   )$ and $\eta \in (1+\lambda, (1+\lambda )^2 )$.

It would be helpful to provide a physical  intuition why  the probabilistic amplifier $\mathcal   Q_g$ works remarkably well so that it
 can achieve our quantum limit.   By acting $Q_N =  \sum_{n=0}^N g ^n \ket{n}\bra{n} $ on  the two-mode squeezed state $\ket{\psi_\xi }= \sqrt{1-\xi^2}\sum_{n=0}^\infty \xi ^n \ket{n}\ket{n} $  we have 
 \cite{Fiurasek09}  
\begin{align}
  Q_N  \ket{\psi_\xi } = \sqrt{1-\xi^2}\sum_{n=0}^{N}  ( g\xi ) ^n \ket{n}\ket{n}  \label{outtm}.
\end{align} This means the resultant (unnormalized) state is proportional to another two-mode squeezed state in the truncated photon-number subspace, i.e.,   $  Q_N  \ket{\psi_\xi } \propto  \ket{\psi_{g \xi} }$. It thus effectively enhances the two-mode squeezed interaction  as $\xi \to g \xi  $ (See section \ref{DistilBound} for a specific statement on the  strength of entanglement).  On the other hand, it has been known that   the two-mode squeezed state minimizes the uncertainty product of Einstein-Podolsky-Rosen-like operators $\ave{\Delta^2 (\hat x_A - g_x \hat  x_B)} \ave{ \Delta^2 (\hat  p_A + g_p \hat p_B)}$ \cite{Namiki13J}. This quantity  appears  in Eq.~\eqref{usr1}, and by  construction its minimum  is responsible for our quantum limit of Eq.~\eqref{AUP2}.  Therefore, we have a simple physical picture that, starting from a two-mode squeezed state $\psi_\xi$, the NLA $\mathcal Q_g $ enables us to produce another  two-mode squeezed state $\psi_{\xi^\prime }$ so as to minimize the corresponding quantum uncertainty (with a certain probability and a finite error).    
This picture would also explain why the NLA $\mathcal Q_g$ could achieve the optimal fidelity in the fidelity-based amplification limit \cite{Chir13}. The optimal fidelity can be related to the maximum eigenvalue of a density operator in the form of $M =\int d^2 \alpha p_\lambda (\alpha ) \ketbra{\alpha }{\alpha  } \otimes  \ketbra{\kappa \alpha^* }{\kappa  \alpha ^* } $ (See Eq.~(15) of \cite{Namiki11R}), and the eigenstate that gives the maximum eigenvalue is a two-mode squeezed state \cite{Namiki11R}.

Now, we can reach the following two statements for the normal amplification/attenuation process:  (i) Our quantum limitation on the amplification/attenuation process  behaves as a tight inequality including the case of the phase-sensitive amplification process; (ii) In order to  demonstrate an advantage of a non-Gaussian amplifier over the Gaussian devices, one needs to operate the amplifier in the regime $\eta \in (1, (1+\lambda )^2 ) $.    
We will address the case of the phase-conjugate amplification/attenuation process in the next subsection.

\subsection{Phase conjugation} \label{ConjugateAmplification}
 Our bound on the uncertainty product in Eq.~\eqref{AUP2} for the phase-conjugate process is equivalent to the bound of the classical limit due to entanglement breaking channels in  Ref.~\cite{Namiki-Azuma13x}. Hence, our bound can be achieved by the following measure-and-prepare scheme
\begin{eqnarray}
\mathcal {A}_G^* (\rho) =  \pi^{-1} \int d^2 \alpha \bra{\alpha}{\rho} \ket{\alpha}\ketbra{\sqrt G {\alpha^*} }{ \sqrt G {\alpha^*}}. 
\end{eqnarray} with $G =  \eta   /(1+ \lambda) ^2$ for the case of  symmetric  gain pair  $\eta = \eta  _x= \eta _p$. For the asymmetric case, 
 the bound can be achieved by  adding the squeezer on the channel as $\mathcal E (\rho) = S \mathcal A_G^* (\rho) S^\dagger$  similar to the flow of Eqs.~\eqref{asgain},~\eqref{acase},~and~(\ref{36}).  
 It concludes the tightness of our quantum limit in Eq.~(\ref{AUP2}) for the case of the phase-conjugation task.

As a summary of this section~\ref{AttainabilityAmplimit}, we have investigated  attainability of our quantum limit given in Eq.~\eqref{AUP2}. For the normal amplification task, it has been shown that there are  two parameter regimes, one that the well-known Gaussian amplifier  achieves  our  quantum limit and the other that a probabilistic non-Gaussian amplifier outperforms the Gaussian amplifier. Specifically, we have shown that the NLA outperforms the Gaussian amplifier and asymptotically achieves our bound  in the  parameter regime $\eta \in (1, (1+ \lambda)^2 )$.  For the phase-conjugation task, our  quantum limit can be  achieved by a Gaussian phase-conjugation channel described by an entanglement breaking map. These structures  repeat the results of the optimal amplification design for the fidelity-based amplification limit given in Ref.~\cite{Chir13}. Hence, it suggests that the optimality of  amplifiers could be addressed straightforwardly by using canonical variables without invoking a fidelity-based figure of merit despite recent studies are more focusing on the property of fidelities \cite{Namiki11R,Chir13,Pande13}.   
 Our results also suggest that  canonical uncertainty relations still play  a significant role in determining quantum limitations on a general physical process.   
 
In the next section we will introduce a different viewpoint on our framework of amplification limits. %

\section{Gaussian amplification limit and entanglement distillation} \label{DistilBound}
In this section, we find an interesting connection between  our amplification  limit and entanglement distillation protocols.  In subsection \ref{Yahoo},  we show  that the no-go theorem for Gaussian entanglement distillation imposes  a physical  limitation on amplifiers composed of Gaussian operations. Then, it turns out that the NLA \cite{Fiurasek09} (the probabilistic amplifier  $\mathcal   Q_g$  of the previous section)
 is actually breaking this limit and regarded as a process of entanglement distillation. In subsection \ref{Yeah},  we show  that our amplification limit, conversely, provides an asymptotically tight limitation on entanglement distillation. This immediately implies that the NLA is an optimal entanglement distillation  process.

\subsection{A tight no-go bound on Gaussian entanglement distillation and a criterion for entanglement distillation by a non-Gaussian amplifier} \label{Yahoo}

Let us define the Einstein-Podolsky-Rosen (EPR) uncertainty for  the density operator $J$ of a two-mode system $AB$ as  \cite{Giedke03}
\begin{eqnarray}
\Delta (J):= \min \left\{ 1, \frac{1}{2} \ave{\Delta^2 (\hat x_A -\hat  x_B)+ \Delta^2 (\hat  p_A +\hat p_B)}_J \right\}.    \label{eprUC} 
\end{eqnarray}  
It determines the entanglement of formation (EOF) for symmetric Gaussian states \cite{Giedke03} and generally gives  a lower bound of  EOF for two-mode states   \cite{{Rig04},{Nicacio13}}, \begin{eqnarray}
E(J) \ge  f[\Delta (J)],      \label{LWD} 
\end{eqnarray}  where $f$ is a decreasing function of $\Delta$ defined  in Ref.~\cite{Giedke03}, and the equality holds when $J$ is a symmetric Gaussian state.  It also suggests that a  smaller EPR uncertainty implies a higher entanglement. 
 Note that Theorem 1 of Ref.~\cite{Rig04} is proven without using the property that the state $\rho$ is a Gaussian state. Hence, the EPR uncertainty gives a lower bound of the EOF not only for two-mode Gaussian states but also for general two-mode states.  
  The EPR uncertainty for the two-mode squeezed state $\ket{\psi_\xi }= \sqrt{1-\xi^2 } \sum_n \xi^n \ket{n}\ket{n}$  can be written as 
\begin{eqnarray}
\Delta (\psi_\xi)=\frac{(1-\xi)^2}{1- \xi^2 }, \label{EPRTMS} 
\end{eqnarray} and the EOF is formally given   by 
\begin{eqnarray}
E(\psi_\xi )= f [\Delta (\psi_\xi)].  \label{EOFTMS}
\end{eqnarray}

Let us consider the case of  $g_x =g_p =1$ in our proof of Eq.~\eqref{AUP2}. Then, with the help of Eqs.~\eqref{LHS}~and~\eqref{paragx}, the EPR uncertainty for a general state $J =  \mathcal E_A  \otimes  I_B  ( \ketbra{\psi_\xi}{\psi_\xi} ) /P_s $  can be associated with the MSDs of Eq.~\eqref{defbarxp} 
 as
\begin{eqnarray}
\Delta (J)&=& \frac{1}{2} \left( \bar V_x^{(\textrm{Prob})}+ \bar V_p^{(\textrm{Prob})} - 1\right) \nonumber \\ 
&=& \bar V^{(\textrm{Prob})} - \frac{1}{2}, \label{avavava}
\end{eqnarray} where $\bar V^{(\textrm{Prob})} = ( \bar V_x ^{(\textrm{Prob})}+   \bar V_p^{(\textrm{Prob})})/2 $ is an average of the MSDs  
 and $   \bar V_z^{(\textrm{Prob})}:= \bar V_z/P_s$.  
When $\mathcal E (\rho ) = \rho $ ($ \mathcal E $ is an identity map),  $J$ is the two-mode squeezed state. Then, substituting  the condition  $g_x =g_p =1$ into Eq.~\eqref{paragx} we have $ \eta = 1 + \lambda = 1/ \xi^2 $.  From this relation and Eq.~\eqref{EPRTMS} we can write  
\begin{eqnarray}
\Delta (\psi_\xi)=\frac{1}{\lambda  }(\sqrt \eta -1 )^2.  \label{monoo} 
\end{eqnarray} 

 Since Gaussian entanglement cannot be distilled by Gaussian local operations and classical communication~\cite{Fiurasek,Giedke2002}, we have
\begin{eqnarray}
E (\psi_\xi) \ge E (J),  \label{monoo0} \end{eqnarray} whenever   $\mathcal E$  is a Gaussian operation.   
Concatenating  Eqs.~ \eqref{LWD},~\eqref{EOFTMS},~and~\eqref{monoo}, 
 we obtain 
\begin{eqnarray}
f  [ \Delta (\psi_\xi)]  \ge f [\Delta(J)].  \label{monoo0} \end{eqnarray} 
Since $f$ is a decreasing function of $\Delta $, this implies  
\begin{eqnarray}
\Delta ( \psi_\xi) \le \Delta (J). \label{KKDA}
\end{eqnarray} 
 This means that  the EPR uncertainty of a two-mode squeezed state cannot be reduced by any  local Gaussian operation. 
 
 Substituting  Eqs.~\eqref{avavava} and  \eqref{monoo} into Eq.~\eqref{KKDA}  we obtain 
\begin{eqnarray} \bar V^{(\textrm{Prob})} \ge  \frac{1}{\lambda  }(\sqrt \eta -1 )^2 + \frac{1}{2}. \label{avava}
\end{eqnarray} 
This is a physical limitation that bounds the average of the MSDs when $\mathcal E$ is a Gaussian CP map. Interestingly, the right-hand side of Eq.~\eqref{avava}  coincides with  the right-hand side of the second equation in Eqs.~\eqref{result1}. Therefore, this bound is tight and achieved by the Gaussian amplifier $\mathcal A_G$ of Eq.~\eqref{eq12}.  It could be helpful to restate this bound in the following form.

\textbf{Theorem 2.--- }
For any Gaussian operation $\mathcal E $ and $\lambda >0$ it holds that 
\begin{eqnarray} \frac{1}{2 P_s}\sum_{z \in \{ x, p \} } \bar V_z (1+ \lambda, \lambda )   \ge  \frac{1}{\lambda  }(\sqrt {1+ \lambda }-1 )^2 + \frac{1}{2},  \label{th2}
\end{eqnarray}  where  $P_s$ and $\bar V_z $ are given by  Eqs.~\eqref{defPS}~and~\eqref{defbarxp},  respectively.   

\textbf{Proof.}--- See   the  above discussion and Eq.~\eqref{avava}.
\hfill$\blacksquare$

Our theorem 2 of  Eq.~\eqref{th2} 
  can be regarded  as an amplification limit for Gaussian operations.  In addition, it {\it per se}
  presents the Gaussian limitation on manipulating   the EPR correlation. Hence, any violation of  Eq.~\eqref{th2} signifies a probabilistic enhancement of entanglement and a non-Gaussian advantage of entanglement distillation.  In other words,  breaking the condition in Eq.~\eqref{th2} is a clear criterion for  an experimental demonstration of entanglement distillation.
Furthermore, such a benchmark can be verified by using  standard homodyne measurements with an input ensemble of coherent states similar to the recently proposed quantum benchmark \cite{Namiki-Azuma13x}.
 
Note that there are different approaches  to characterize non-Gaussian entanglement generation  \cite{Kitagawa06,Navarre12,Bart13}. 
Our result here is directly determined by the no-go theorem for Gaussian entanglement distillation and applicable to local filtering operations acting on a single mode. Moreover, it ensures an enhancement of the EOF.  It would be valuable to investigate how one can beat our boundary of theorem 2 by using the state of the art technology in  photonic quantum state engineering \cite{Wenger04,Zava04,Parigi07,Ourjoumtsev07,Zava08,Kim08,Takahashi08,Takahashi10,Name10,Neergaard-Nielsen2013}  and whether the experimental demonstrations of probabilistic amplifications \cite{Ferreyrol2010,Usuga2010,Zavatta2011} can fulfill our criterion.   

Although Eq.~\eqref{th2} gives a tight limitation for Gaussian operations, our statement is severely restricted for the single point $\eta =1+\lambda $ of the curve achieved by the Gaussian channel in  the second inequality of Eq.~\eqref{result1} (See FIG.~\ref{fig:ampfig1.eps}b). Therefore, it remains open how to determine such an amplification limit on  the class of Gaussian operations  for the entire parameter space $\eta \in (1, (1+\lambda)^2 )$. 

\subsection{Amplification limit as a physical limit on distillation of entanglement via local filtering operations}\label{Yeah}

 We show our amplification limit of Eq.~\eqref{AUP2}  presents a bound for  minimizing the EPR uncertainty when one uses the local filtering operation described by a stochastic quantum channel.

In contrast to our distillation bound  for Gaussian operations in  Eq.~\eqref{th2}  we have the following statement for general CP maps: 

\textbf{Corollary.--- }
For any  operation $\mathcal E  $ and $\lambda >0$ it holds that 
\begin{eqnarray} \frac{1}{P_s}\sum_{z \in \{ x, p \} } \bar V_z (1+ \lambda, \lambda )   \ge  1 ,  \label{th3}
\end{eqnarray}   where  $P_s$ and $\bar V_z $ are given by  Eqs.~\eqref{defPS}~and~\eqref{defbarxp},  respectively.

\textbf{Proof.}---Recalling $\eta = 1+ \lambda$ and $  \bar V_z^{(\textrm{Prob})} - 1/2 \ge 0$ in Eq.~\eqref{avavava} we can show that 
\begin{eqnarray}
\Delta (J)&=& \frac{1}{2} \left( \bar V_x^{(\textrm{Prob})}+ \bar V_p^{(\textrm{Prob})} - 1\right) \nonumber \\ 
&\ge & \sqrt{ \left( \bar V_x^{(\textrm{Prob})} - \frac{1}{2} \right)  \left( \bar V_p^{(\textrm{Prob})} - \frac{1}{2} \right)}  \ge 0,  \label{DLimit}
\end{eqnarray} where we use the relation $a + b \ge  2 \sqrt{ a b}$ for $\{ a,b\} \ge 0 $ and our theorem 1 of Eq.~\eqref{AUP2}. This proves  Eq.~\eqref{th3}. \hfill$\blacksquare$ 

The property of $\Delta (J) \ge 0$ itself can be obtained from the definition of the EPR uncertainty in Eq.~\eqref{eprUC}, and this Corollary is rather trivial. An interesting point here is that the minimum of this inequality, which is the bound on the entanglement distillation process starting from a two-mode squeezed state,  is asymptotically achievable by the probabilistic amplifier $\mathcal Q_g$  in Eq.~\eqref{alphaW}.  Hence, the NLA  is not only a probabilistic  amplifier that enables us to break the no-go bound on Gaussian operations in Eq.~\eqref{th2}, but also provides an optimal process that asymptotically achieves the physical limitation  of Eq.~\eqref{th3}.    
Again, the simple physical picture that, starting from a two-mode squeezed state $\psi_\xi$, the NLA $\mathcal Q_g $ enables us to produce another  two-mode squeezed state $\psi_{\xi^\prime }$, would explain why  this process could be optimum [See Eq.~\eqref{outtm}].  It would be worth noting that a quantum benchmark inequality (Corollary 1 of \cite{Namiki-Azuma13x}) with $\eta = 1+\lambda $ corresponds to 
\begin{eqnarray} \frac{1}{P_s}\sum_{z \in \{ x, p \} } \bar V_z (1+ \lambda, \lambda )   \ge  3   \label{th3-}.
\end{eqnarray}  The equality implies  $\Delta (J) = ( \bar V_x^{(\textrm{Prob})}+ \bar V_p^{(\textrm{Prob})} - 1)/2 =1$. Hence, the separable point $E(J)=0$ is consistent with the entanglement breaking limit.

In this section~\ref{DistilBound}, we have found an insightful  interrelation between our amplification limit and continuous-variable entanglement. It has been shown that the no-go theorem for Gaussian entanglement distillation gives a limitation on Gaussian amplifiers. Thereby, we have pointed out that the NLA can break this limit and would be useful to demonstrate  a significance of  non-Gaussian process. In addition, it  turned out  that  our amplification limit determines a physical limitation of  entanglement distillation due to local filtering operations. 
Note that one can find different links between probabilistic amplifiers and entanglement distillation in Refs.~\cite{Xiang2010a,Fiurasek2010,Chrzanowski2014}. Note also that local photon-subtraction and addition could  reduce the EPR uncertainty, and enhance entanglement  \cite{Lee11}.

\section{conclusion and remarks}\label{ConcRemark}
In this paper we have presented an uncertainty-product form of quantum amplification limits based on the input ensemble of Gaussian distributed coherent states, and successfully revived the key role of canonical uncertainty relations in determining a general quantum limit.  Our amplification limit retrieves basic properties of  the traditional amplification  limit  without assuming the linearity condition. Moreover, it is usable for general stochastic quantum channels, hence  probabilistic amplifiers.   Given a physical process one can test how close the performance of the process approaches to the ultimate quantum limit  via an accessible input set of coherent states  and standard homodyne measurements. We have also identified the parameter regime where Gaussian channels cannot achieve our bound but the NLA \cite{Ralph09} asymptotically achieves our bound. 
 In addition, we have derived an  amplification limit on  Gaussian operations  by using the no-go theorem for Gaussian entanglement distillation. This in turn shows that beating this limit implies a  clear advantage of non-Gaussian processes in reducing the EPR uncertainty,  and  establishes a simple criterion for  entanglement distillation.  Thereby, we have found that the NLA  is not only an amplifier whose action is  useful for an enhancement of entanglement but also constitutes an optimal local filtering process for reducing the EPR uncertainty.  
   It would be valuable to investigate how one can demonstrate such a  non-Gaussian advantage 
    by using the state of the art technology in  photonic quantum state engineering \cite{Wenger04,Zava04,Parigi07,Ourjoumtsev07,Zava08,Kim08,Takahashi08,Takahashi10,Name10,Neergaard-Nielsen2013} as well as in the experiments of the noiseless amplification \cite{Xiang2010a,Chrzanowski2014,Ferreyrol2010,Usuga2010,Zavatta2011}.

Unfortunately, our result on  the Gaussian amplification limit works for a rather restricted set of the parameters.  The possibility to extend Theorem~2 beyond the present constraints is left for future works.  
It  remains open whether  (i) a probabilistic Gaussian channel might outperform the deterministic Gaussian channel  and (ii) Gaussian channel could be  an optimal  trace-preserving map (both regarding the parameter regime $\eta \simeq 1 + \lambda $).  The second statement (ii) is true for the case of the fidelity-based amplification limit  \cite{Chir13}, while the validity of the first statement (i) is unclear. 
 It is also open whether  (iii) one can signify the non-Gaussian advantage on entanglement distillation from the viewpoint of the fidelity-based approach.

\acknowledgments
RN was supported by the DARPA Quiness program under prime Contract No. W31P4Q-12-1- 0017 and Industry Canada. 
This work was partly supported by the Grant-in-Aid for the
Global COE Program ``The Next Generation of Physics, Spun from Universality and Emergence''  from the Ministry of Education, Culture, Sports, Science and Technology of Japan (MEXT).

\appendix

\section{Connection to the amplification uncertainty principle} \label{AP1}
In this appendix we show that our amplification limit (for the case of the uniform distribution $\lambda \to 0 $) coincides with the familiar traditional form of   amplification limits given in Ref.~\cite{amp}.  

Let us recall the \textit{amplifier uncertainty principle} (AUP) in Ref.~\cite{amp}. We consider linear transformation of a single mode field so that the first moments are linearly amplified
  with possibly phase depending gain factor $(G_x, G_p)$ as 
\begin{eqnarray}
\ave{\hat Y_x}= \sqrt{ G_x}  \ave{\hat X_x}, \   \ave{\Delta ^2 \hat Y_x}  =   G_x  \ave{\Delta ^2  \hat X_x}   +  \mathcal N_x,   \nonumber \\ 
\ave{ \hat Y_p}= \pm \sqrt{ G_p}  \ave{\hat  X_p}, \  \ave{\Delta ^2 \hat  Y_p} =   G_p \ave{\Delta ^2 \hat  X_p}  +  \mathcal N_p ,  \label{Lin}
\end{eqnarray} where $\hat X$ and $\hat Y$ denote input and output quadratures, respectively. They satisfy the canonical commutation relation $[\hat X_x, \hat X_p ] = [\hat Y_x, \hat Y_p ] = i  $. 
 The upper sign and lower signs in Eq.~\eqref{Lin} respectively indicate the cases of the normal amplification/attenuation process and the phase-conjugation process.   We may  focus on the property of added noise terms: 
\begin{eqnarray} \mathcal N_x &= \ave{\Delta ^2 \hat Y_x}  -G_x    { \ave{\Delta ^2  \hat X_x}}  , \nonumber \\  \mathcal N_p& =  {\ave{\Delta ^2 \hat Y_p}}  - G_p    { \ave{\Delta ^2  \hat X_p}} . \label{adnoise}\end{eqnarray} It 
 tells us an amount of additional noise imposed by the channel because the second terms in Eqs.~(\ref{adnoise}) represent  the variance of an input state. 
The AUP gives a physical limit for  CP trace-preserving maps satisfying Eq.~(\ref{Lin}):  
\begin{eqnarray}
 \label{AUR0} 
\mathcal N _x \mathcal N _p \ge \frac{1}{4} \left| \sqrt{G_x G_p} \mp  1 \right|^2.
\end{eqnarray}   
Note that  in Ref. \cite{amp} the AUP is defined through the added noise number 
 $A_z  : =   \mathcal N_z  / G_z  =  \ave{\Delta ^2 \hat Y_z}/G_z -  \ave{ \Delta ^2 \hat X_z}  $.

In order to link Eq.~(\ref{AUR0}) to our amplification limit in Eq.~(\ref{AUP2}), we consider 
the input of coherent states  $\rho_\alpha = \ketbra{\alpha}{\alpha }$ with the shorthand notation of Eq.~\eqref{ShNote}. It implies  
 \begin{eqnarray}
\ave{\Delta ^2 \hat X_x} = \ave{\Delta ^2 \hat X_p} =1/2,  \label{Lin0} \\
\ave{\hat X_x}=   x_\alpha,\ \ave{ \hat X_p}=   p _\alpha.  \label{Lin2} 
\end{eqnarray} 
Using Eqs.~\eqref{Lin} and \eqref{Lin2} we can write  
\begin{align}
\ave{\Delta^2 \hat Y_x}&=& \ave{ \hat Y_x ^2 } - \ave{  \hat Y_x}^2 =  \tr \left[  ( \hat x - \sqrt{G_x} x_\alpha)^2  \mathcal E (\rho_\alpha )  \right]  
,   \nonumber \\ 
\ave{\Delta^2 \hat Y_p}&=& \ave{ \hat Y_p ^2 } - \ave{  \hat Y_p}^2 =  \tr \left[  (\hat p  \mp  \sqrt{G_p} p_\alpha)^2    \mathcal E (\rho_\alpha )  \right]. \label{lin3}  
\end{align} 
Due to the linearity assumption, we can write any average of the variance $(\ave{\Delta^2 \hat Y_x},\ave{\Delta^2 \hat Y_p})$ over the coherent-state amplitude $\alpha $ as  the variances for a single coherent state.  Hence, it holds that \begin{eqnarray}
\int p_\lambda (\alpha )  (\ave{\Delta^2 \hat Y_x},\ave{\Delta^2 \hat Y_p}) d^2 \alpha = (\ave{\Delta^2 \hat Y_x},\ave{\Delta^2 \hat Y_p}) \label{A9}  . 
\end{eqnarray} 

Concatenating  Eqs.~\eqref{adnoise},~\eqref{Lin0},  \eqref{lin3}, and  \eqref{A9}  we can write  
\begin{align} \mathcal N_x &= \underbrace{ \int p_\lambda ( \alpha) \tr \left[  ( \hat x - \sqrt{G_x} x_\alpha)^2  \mathcal E (\rho_\alpha )  \right] d^2 \alpha}_{\bar V_x (G_x, \lambda )}   - G_x /2, \nonumber \\ 
 \mathcal N_p&   =  \underbrace{\int p_\lambda ( \alpha)  \tr \left[  (\hat p  \mp  \sqrt{G_p} p_\alpha)^2    \mathcal E (\rho_\alpha )  \right]  d^2 \alpha}_{\bar V_p (G_p, \lambda )}   -G_p/2 , \label{adnoise2}\end{align}  
where the underbracing terms, $\bar V_x$  and $\bar V_p$,  come from Eq.~\eqref{defbarxp}.  
Substituting Eqs.~\eqref{adnoise2}  into Eq.~\eqref{AUR0} we can re-express the AUP as  
\begin{eqnarray}
 \label{AUR00} 
\prod_{z = x, p } \left[  \bar V_z   (G_z, \lambda ) -  {G_z}/{2 } \right]   \ge \frac{1}{4} \left| \sqrt{G_x G_p} \mp  1 \right|^2. 
\end{eqnarray} 
It would be instructive to illustrate the gain-dependence for symmetric cases as in FIG.~\ref{fig:ampfig1.eps}(b).
For the normal amplification process with 
$G = G_x = G_p$ and $\bar V = \bar V_x = \bar V_p $ we have 
\begin{eqnarray}
 \label{pil} 
   \bar V   \ge \frac{1}{2} ( G + \left| G  - 1 \right|) . 
\end{eqnarray} 
Similarly,  for the phase-conjugation process,  we have %
\begin{eqnarray}
 \label{pil2} 
   \bar V   \ge G +  \frac{1}{2}    . 
\end{eqnarray} 
We thus  apparently observe that  the structures of Eqs.~\eqref{pil}~and~\eqref{pil2}  are the same as those of
 Eqs.~\eqref{eeto}~and~\eqref{eeeto}, respectively. 

On the other hand,  substituting  $\{\lambda, \eta_x, \eta_p  \} = \{0, G_x,  G_p  \} $ in Eq.~\eqref{AUP2} and assuming $\mathcal E$ is a CP trace-preserving map  we can write our amplification limit as  
\begin{eqnarray}\prod_{z = x, p } \left[  \bar V_z   (G_z, 0) -  {G_z}/{2 } \right] 
 \ge \frac{1}{4}\left|   { \sqrt{G_x G_p}  } \mp 1  \right|^2 \label{AUP2d}.  \end{eqnarray}
Comparing  this relation with  Eq.~\eqref{AUR00} 
 we can see that our amplification limit  coincides with the AUP  in the limit of $\lambda \to 0 $. 
It is clear  from FIG.~\ref{fig:ampfig1.eps}(b) that the inequalities of  Eq.~\eqref{pil} [Eq.~\eqref{pil2}] can be violated  for any finite width of the distribution $\lambda > 0$ whenever $\eta >1 $ [$\eta>0$].

\end{document}